\documentclass[runningheads]{llncs}
\usepackage{amsmath,amssymb,amsfonts}
\usepackage{algorithmic}
\usepackage{graphicx}
\usepackage{textcomp}
\usepackage{xcolor}
\usepackage{booktabs}
\usepackage{graphicx}
\usepackage{stfloats}
\usepackage{subcaption}
\usepackage{multirow}
\usepackage{url}

\usepackage{makecell}
\usepackage{float}
\usepackage{cite}
\usepackage{hyperref}

\begin{document}
\title{How Secure is Your Website? A Comprehensive Investigation on CAPTCHA Providers and Solving Services}
\author{Rui Jin\inst{1,2}\thanks{First Author and Second Author contribute equally to this work.} \and
Lin Huang\inst{1,2} \and
Jikang Duan\inst{1,2} \and
Wei ZHAO\inst{3,4}  \and
Yong Liao\inst{1,2} \and
Pengyuan Zhou\inst{1,2}}

\institute{University of Science and Technology of China \and
Research Center For Data to Cyberspace \and
Anhui Engineering Research Center for Intelligent Applications and Security of Industrial Internet \and
Anhui University of Technology} 

\maketitle

\begin{abstract}
Completely Automated Public Turing Test To Tell Computers and Humans Apart (CAPTCHA) has been implemented on many websites to identify between harmful automated bots and legitimate users. However, the revenue generated by the bots has turned circumventing CAPTCHAs into a lucrative business. Although earlier studies provided information about text-based CAPTCHAs and the associated CAPTCHA-solving services, a lot has changed in the past decade regarding content, suppliers, and solvers of CAPTCHA.  We have conducted a comprehensive investigation of the latest third-party CAPTCHA providers and CAPTCHA-solving services' attacks. We dug into the details of CAPTCHA-As-a-Service and the latest CAPTCHA-solving services and carried out adversarial experiments on CAPTCHAs and CAPTCHA solvers. The experiment results show a worrying fact: most latest CAPT-CHAs are vulnerable to both human solvers and automated solvers. New CAPTCHAs based on hard AI problems and behavior analysis are needed to stop CAPTCHA solvers. 
\end{abstract}

\section{Introduction}

Completely Automated Public Turing Test To Tell Computers and Humans Apart, or CAPTCHA, has been widely deployed on websites to fight against these malicious activities. As the name suggests, CAPTCHAs are puzzles that are difficult for programs to solve but can be easily solved by humans. The primitive and most common CAPTCHAs are images that contain distorted and blurred letters. Users need to correctly recognize these letters. The arms race between CAPTCHAs and CAPTCHA solver algorithms has made them more and more complicated. Nowadays, many popular websites turn to more difficult CAPTCHAs, e.g., requiring users to select one image that contains certain animals from several images or listen to audio of several words with background noise and spell them.

The websites risk losing annoyed users and insist on deploying CAPTCHA since it helps prevent automated bots from carrying out malicious activities such as spamming, brute-force attacks, data scraping, etc. Requiring users to solve a CAPTCHA adds an extra layer of security by verifying that the user is human. This further help prevents automated systems from manipulating or distorting data, such as voting systems, online polls, or user-generated content, so that the accuracy and reliability of data can be maintained. Last, CAPTCHAs help prevent fraudulent activities, such as account creation or online transactions carried out by automated bots. Adding a verification step reduces the risk of fraudulent activities and enhances the overall security of online platforms.

Widely deployed CAPTCHAs have made solving CAPTCHAs a profitable business in turn. There are two methods to solve CAPTCHAs at a low cost: advanced AI or outsourced cheap human labor. The former method is commonly used to solve relatively old CAPTCHAs that can no longer stump the latest AIs. Since some small/old websites haven't updated their outdated text-based CAPTCHAs, this method still occurs in most CAPTCHA-solving services retailers. The latter method is used when no known AI in the community can achieve a high success rate against the target CAPTCHA. Retailers usually hire workers from low-income areas to reduce costs. Since CAPTCHAs are designed to be easy for humans, the entry barrier for this job is very low. The price of solving CAPTCHAs using manpower can be as low as several dollars per thousand CAPTCHAs.

The development of AI has increased the difficulty of developing CAPTCHAs. As CAPTCHA-As-a-Service becomes mainstream, great changes have taken place in CAPTCHA providers and CAPTCHA solvers. However, few studies have demonstrated the current situation of the war between CAPTCHA providers and underground CAPTCHA-solving services. In this paper, we present the mechanic behind third-party CAPTCHA providers and CAPTCHA-solving services. Furthermore, we test the CAPTCHA-solving services against popular CAPTCHA providers to unveil more details of underground CAPTCHA-solving services. Our contributions include:
\begin{itemize}
\item[-] Investigating and summarizing CAPTCHA-As-a-Service.
\item[-] Unveiling the attack details and service quality of underground CAPTCHA-solving services. 
\end{itemize}

The structure of this paper is as follows. In Section ~\ref{background}, the related background of CAPTCHAs will be introduced. The CAPTCHA-As-a-Service framework and popular CAPTCHA providers will be presented in Section ~\ref{provider}. Section ~\ref{solver} introduces the attack framework and some CAPTCHA-solving services retailers. The adversarial experiments between selected CAPTCHA providers and CAPTCHA-solving services will be shown in Section ~\ref{experiment}. Section ~\ref{conclusion} is the conclusion.
\section{Background}\label{background}
Luis von Ahn et al. proposed the term CAPTCHA in 2003 \cite{von2003captcha}. In this paper, they defined CAPTCHA as a program that can generate and grade tests that most humans can easily pass but current computer programs can't. They also proposed the core of CAPTCHA: challenging AI problems. The design of CAPTCHAs is similar to modern cryptography: cryptography algorithms are based on challenging mathematic problems that can't be solved at a realistic cost, and CAPTCHAs are based on difficult AI problems that the AI community hasn't found a method to solve with high success rate. For example, when they formalized the text CAPTCHAs, they assumed that programs couldn't achieve high accuracy in transformed letter recognition with the available technology. With modern cryptography, we haven't heard of many leakage incidents due to the decryption of encryption algorithms. However, even though many websites have armed themselves with SOTA CAPTCHAs, fake accounts and scalpers haven't disappeared and have become even more rampant. On one hand, the development of AI has made it possible to break some CAPTCHAs with a high success rate. On the other hand, the underground CAPTCHA-solving services using cheap manpower provided a guarantee to solve those CAPTCHAs that no software could pass.

In 2010, Marti Motoyama conducted a detailed investigation of CAPTCHAs and CAPTCHA solvers\cite{motoyama2010re}. Before 2010, the mainstream CAPTCHAs were text-based, and many websites deployed their own CAPTCHA. In this study, the author tested 8 CAPTCHA-solving services against 25 popular websites' CAPTCHAs. The result showed that the services could achieve a correct rate of over 70\% on most websites within 20 seconds. The author also argued that defenders were winning the war against low-cost-effectiveness automated software solvers, which usually cost hundreds or even thousands of dollars yet couldn't achieve promising results.

Most text-based CAPTCHAs were broken using software by 2014 \cite{guerar2021gotta}. Twisted texts can no longer stop software solvers while keeping themselves easy for humans to recognize. Strictly speaking, text-based CAPTCHAs are included in image-based CAPTCHAs: text-based CAPTCHAs ask users to recognize texts from images, while the recognition target of image-based CAPTCHAs can be more varied. In 2012, using photographs from Google Street View, Google reCAPTCHA developed reCAPTCHA v2, which requires the user to identify crosswalks, bikes, buses, etc. Despite the rapid development of semantic computer vision, the attacker must train the model for each new recognition target, significantly increasing the cost of software CAPTCHA solvers. So far, image-based CAPTCHAs are still mainstream. In 2019, Weng et al also tested CAPTCHA-solving services on common image-based CAPTCHAs \cite{weng2019towards}. 152 CAPTCHA-solving services distributed worldwide were confirmed. Most tested CAPTCHA-solving services' success rate against image-based CAPTCHAs ranged from 0.8 to 0.95.

Apart from the content of CAPTCHAs, CAPTCHA providers have also changed in the last decade. Popular websites no longer develop their own CAPT-CHA and instead turn to third-party CAPTCHA providers, e.g., Google reCAPTCHA, Arkose Labs, etc. Typically, third-party CAPTCHAs are implemented in an independent iframe using scripts developed by these providers. After solving the CAPTCHA, the user will receive a token from the CAPTCHA provider. The user can pass the CAPTCHA by sending this token to the website. Since the difficulty of CAPTCHAs varies between providers, CAPTCHA solvers usually classify target CAPTCHAs by providers and set the price respectively.

\section{CAPTCHA providers}\label{provider}
The arms race between CAPTCHAs and CAPTCHA solvers has pushed CAPT-CHAs to become more complicated. Since AIs can easily pass self-developed text-based CAPTCHAs, which have been thoroughly researched, many websites use CAPTCHAs from professional third-party CAPTCHA providers. Few studies have presented the mechanic of third-party CAPTCHAs. In this section, we will present the framework of third-party CAPTCHAs and introduce four third-party CAPTCHA providers: Google reCAPTCHA, Arkose Labs, Geetest, and hCaptcha, since most CAPTCHA-solving services retailers support them, and we will use them as testbeds.

\subsection{Third-party CAPTCHAs framework}
\begin{figure}[h]
\includegraphics[width=\textwidth]{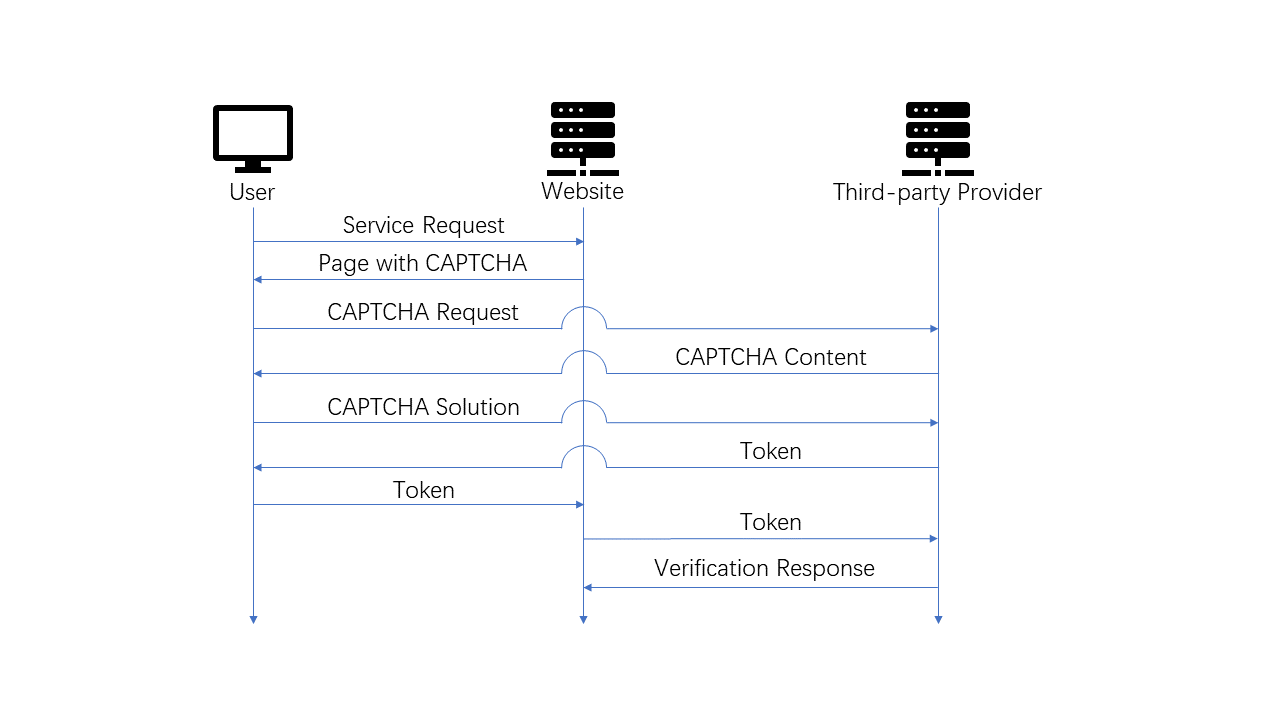}
\caption{Third-party CAPTCHAs Framework} \label{framework}
\end{figure}

Before deploying third-party CAPTCHAs, the website will need to apply a pair of keys from the provider: a public site key and a private secret key. On the client side, most websites will place the CAPTCHA and the public site key in an iframe. As the first line of defense, users usually need to click a check box to request and load the content of the CAPTCHA. The public site key is used here to invoke the CAPTCHA. After solving the CAPTCHA, the client will send the collected data to the provider. If the provider believes the client is a human, a response token will be generated and sent to the client. The client needs to send the response token to the server then. The server can accomplish the verification by consulting the CAPTCHA provider with the received response and the private secret key. This framework is presented in Fig.~\ref{framework}.

\subsection{Popular providers}
\subsubsection{Google reCAPTCHA}
\begin{figure*}[h]
\centering
\begin{subfigure}[b]{0.45\textwidth}
\centering
\includegraphics[height=0.05\textheight]{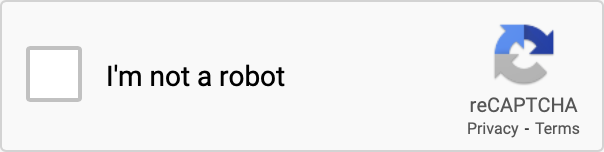}
\end{subfigure}
\begin{subfigure}[b]{0.45\textwidth}
\centering
\includegraphics[height=0.21\textheight]{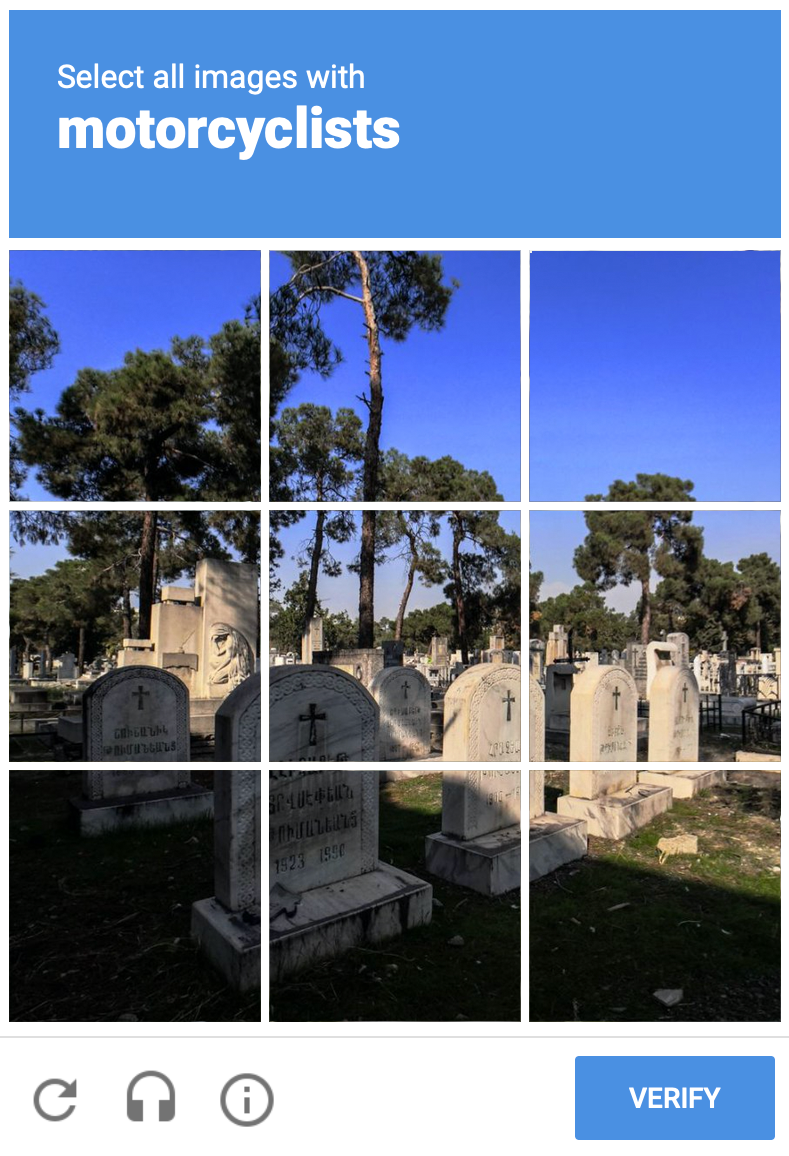}
\end{subfigure}
\caption{Google reCAPTCHA\protect\footnotemark}
\label{Google reCAPTCHA examples}
\end{figure*}

There are three reCAPTCHA versions: reCAPTCHA v1, v2, and v3. reCAPTCHA v1 is text-based and has been shut down since 2018. There are two types of popular reCAPTCHA v2. The first type asks the user to click the “I'm not a robot” checkbox and select images with a certain object from nine candidate images (Fig.~\ref{Google reCAPTCHA examples}). The second type, reCAPTCHA V2 Invisible, doesn't need the user to click the checkbox but requires it to be bonded to a button or invoked programmatically instead. Also, it analysis the user's IP addresses, cookies, mouse movements, etc, to decide the risk and whether it needs to send further challenges to the user. The latest reCAPTCHA v3 is similar to reCAPTCHA V2 Invisible, except it won't send any explicit challenge and will score the user instead of simply classifying the user into human or machine.
\footnotetext{https://anti-captcha.com/apidoc/task-types/RecaptchaV2TaskProxyless}

\subsubsection{GeeTest}
\begin{figure*}[h]
\centering
\begin{subfigure}[b]{0.45\textwidth}
\centering
\includegraphics[height=0.21\textheight]{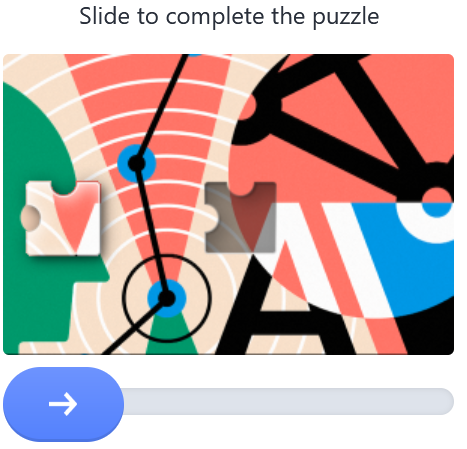}
\end{subfigure}
\begin{subfigure}[b]{0.45\textwidth}
\centering
\includegraphics[height=0.21\textheight]{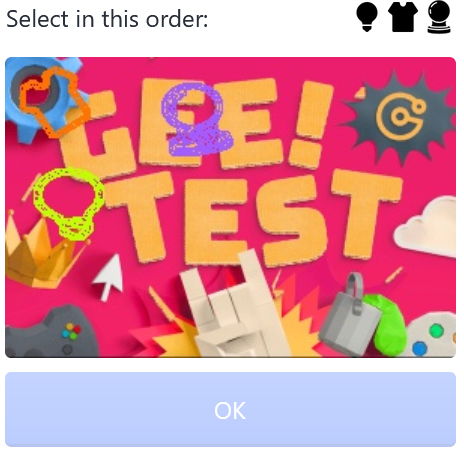}
\end{subfigure}
\begin{subfigure}[b]{0.45\textwidth}
\centering
\includegraphics[height=0.21\textheight]{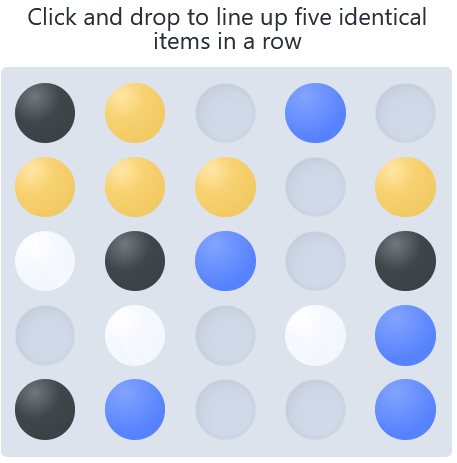}
\end{subfigure}
\caption{GeeTest CAPTCHAs\protect\footnotemark}
\label{GeeTest examples}
\end{figure*}

As shown in Fig.~\ref{GeeTest examples}, GeeTest's feature is its various types of interaction. It might ask the user to slide a piece of puzzle, click icons in a certain order, or swap items to line up identical items in a row. The reason they developed these challenges, they claim, is to collect the user's behavior data so that they can verify the user without accessing other client-side private information.
\footnotetext{https://www.geetest.com/adaptive-captcha-demo}

\subsubsection{FunCaptcha}
\begin{figure*}[h]
\centering
\begin{subfigure}[b]{0.45\textwidth}
\centering
\includegraphics[height=0.21\textheight]{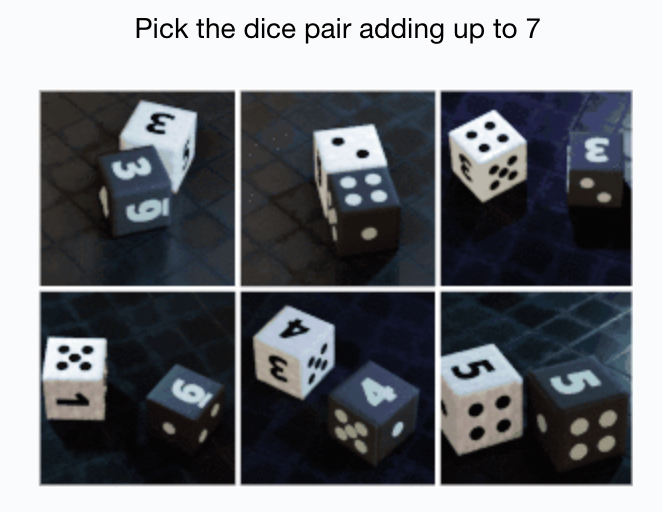}
\end{subfigure}
\begin{subfigure}[b]{0.45\textwidth}
\centering
\includegraphics[height=0.21\textheight]{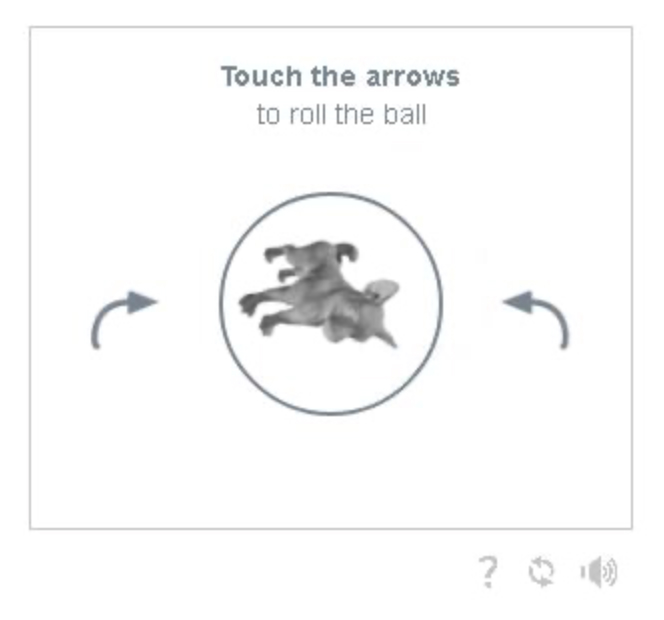}
\end{subfigure}
\begin{subfigure}[b]{0.45\textwidth}
\centering
\includegraphics[height=0.21\textheight]{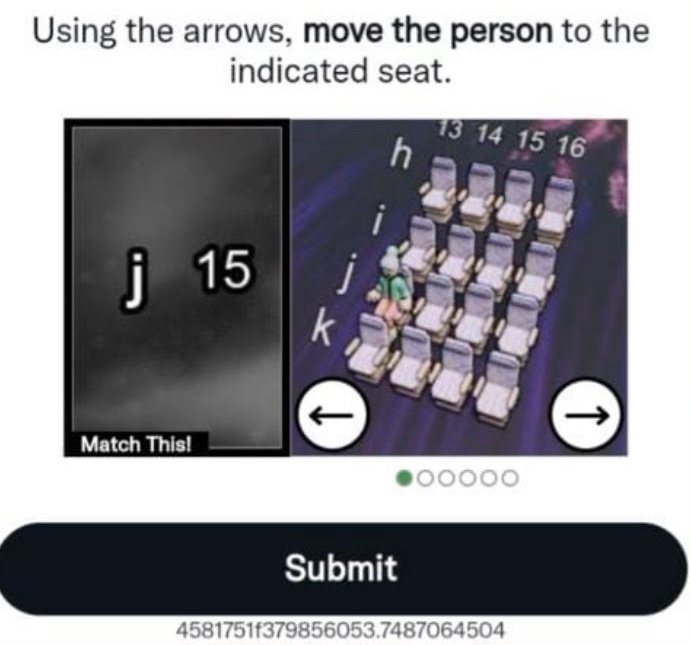}
\end{subfigure}
\begin{subfigure}[b]{0.45\textwidth}
\centering
\includegraphics[height=0.21\textheight]{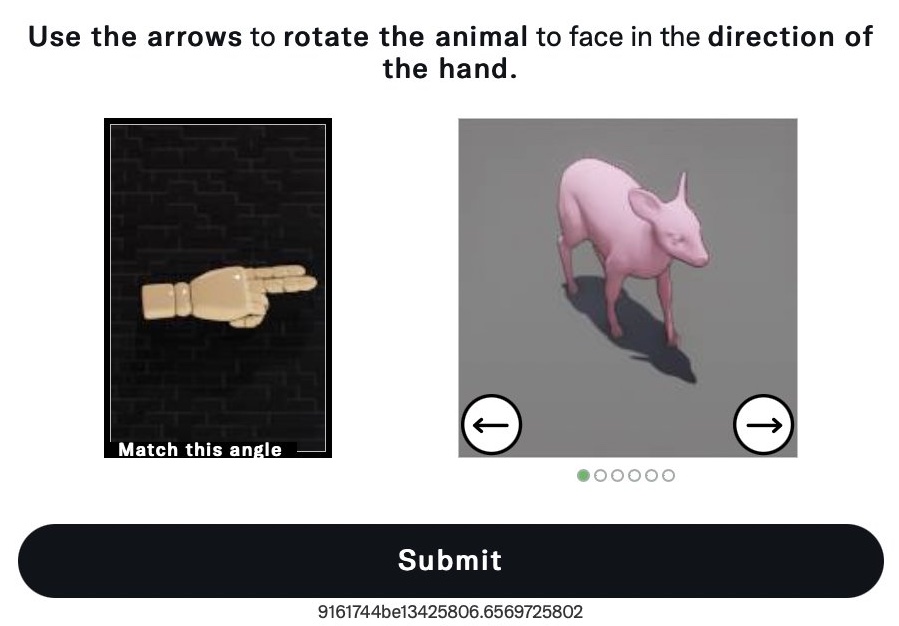}
\end{subfigure}
\caption{FunCaptcha CAPTCHAs\protect\footnotemark}
\label{FunCaptcha examples}
\end{figure*}

FunCaptcha is developed by Arkose Labs. From the example on their website, we can tell available environment information for CAPTCHAs includes but is not limited to the operating system version, the browser version, IP address, and device fingerprints. The challenge contents of FunCaptcha are presented in Fig.~\ref{FunCaptcha examples}.
\footnotetext{https://anti-captcha.com/apidoc/task-types/FunCaptchaTaskProxyless}

\subsubsection{hCaptcha}
\begin{figure*}[h]
\centering
\begin{subfigure}[b]{0.45\textwidth}
\centering
\includegraphics[height=0.05\textheight]{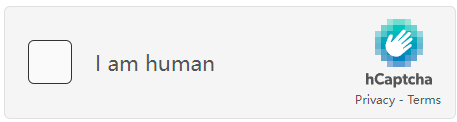}
\end{subfigure}
\begin{subfigure}[b]{0.45\textwidth}
\centering
\includegraphics[height=0.21\textheight]{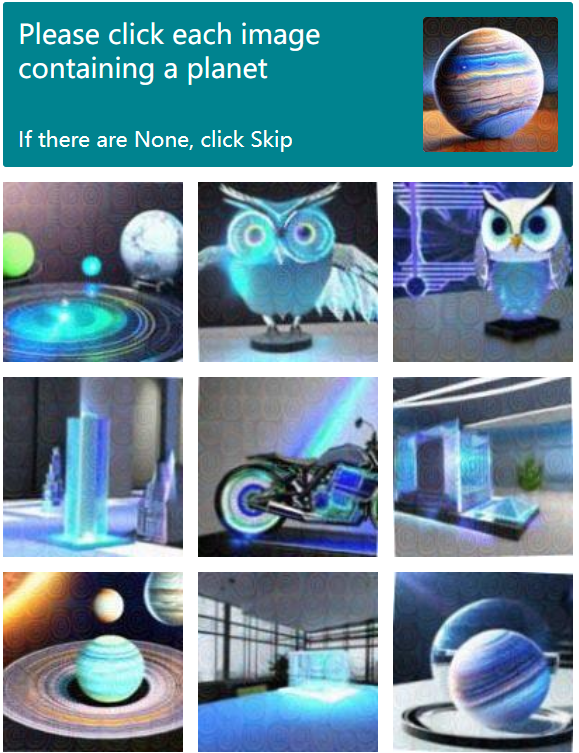}
\end{subfigure}
\caption{hCAPTCHA\protect\footnotemark}
\label{hCAPTCHA examples}
\end{figure*}

hCaptcha (Fig.~\ref{hCAPTCHA examples}) is very similar to Google reCAPTCHA. They claim all reCAPTCHA V2 and V3 features and advertise that the websites can switch from reCAPTCHA to hCaptcha with minimal effort. Another feature is that the difficulty of hCaptcha, which decides whether an explicit challenge will appear and its difficulty, can be manually set.
\footnotetext{https://2captcha.com/demo/hcaptcha?difficulty=moderate}

In summary, we can see two different development paths for CAPTCHAs. The first path is to continue designing new tests using hard AI problems. For example, apart from recognizing objects, FunCaptcha's new challenges require logical thinking and spatial imagination, thus being more difficult for AI. Another path is to collect the user's behavior and environment data and analyze it to classify or score the user. Despite this method has deviated from the original definition and concept of CAPTCHA, Google reCAPTCHA and many other CAPTCHA providers are digging deeper into it and leaving the broken challenges as a way to collect behavior data rather than an effective defense line.
\section{CAPTCHA solvers}\label{solver}
CAPTCHA-solving services retailers follow CAPTCHA providers closely to ensure they support new CAPTCHAs with high demand. Thus, modern retailers differ vastly in attack framework and pricing method from old retailers. In this section, we will introduce some service retailers and how attackers utilize CAPTCHA-solving services to pass third-party CAPTCHAs.

\subsection{Attack framework}
\begin{figure}[h]
\includegraphics[width=\textwidth]{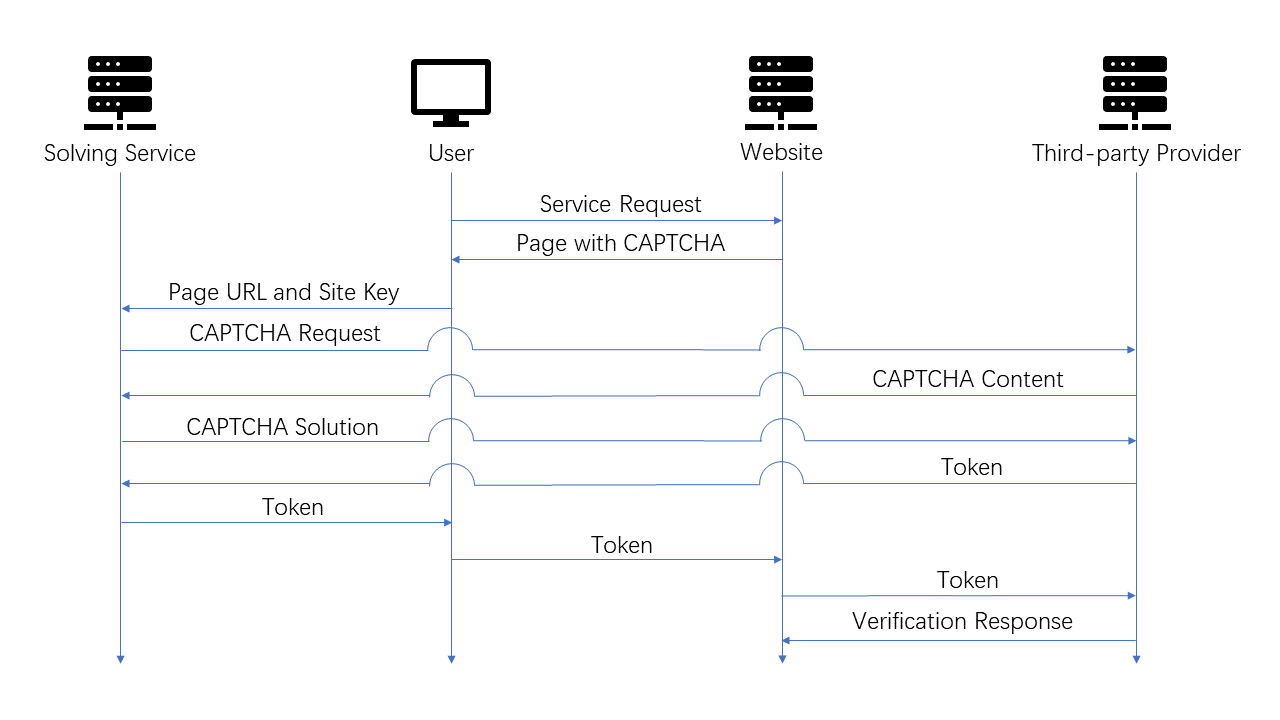}
\caption{Third-party CAPTCHAs Attack Framework} \label{attack_framework}
\end{figure}

The attack method differs for text-based CAPTCHAs and image-based CAPT-CHAs since only recognized texts are needed to solve a text-based CAPTCHA, but the action needed to solve an image-based CAPTCHA varies. Also, the information in a single image might not be sufficient for some CAPTCHAs, e.g., Google reCAPTCHA might replace the clicked image with a new one and ask the user to continue recognizing it. For text-based CAPTCHAs, the attacker only needs to send the image with text, wait for the recognized texts, then put them into the input box. The attack framework for image-based CAPTCHAs is represented in Fig.~\ref{attack_framework}. The user will not interact with the CAPTCHA. Instead, the user must extract information from the page that the solving services require to load the CAPTCHA. Then, the solving services will pretend to be the user and obtain the response token. After the user receives it, the user only needs to accomplish the remaining communication process after passing the challenge, which is usually filling the token in an element on the page and submitting the form, or 
dispatching an event with it.

\subsection{Service retailers}
\begin{figure}[h]
\includegraphics[width=\textwidth]{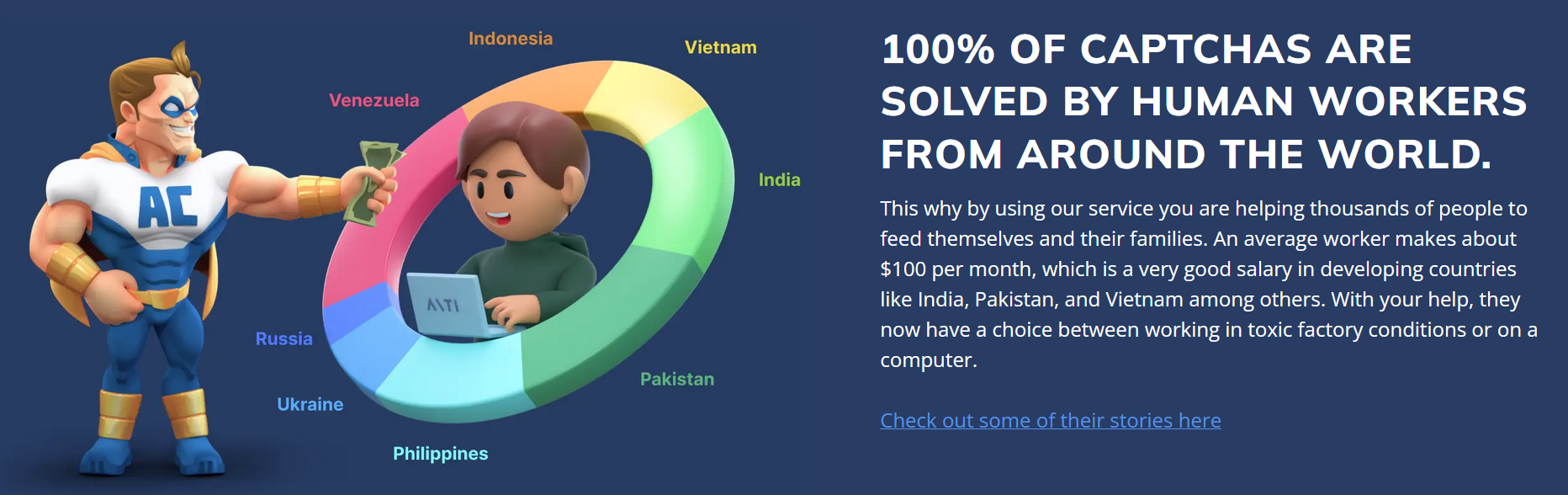}
\caption{AntiCaptcha advertising all CAPTCHAs are solved by humans distributed in developing countries\protect\footnotemark.} \label{advertisement}
\end{figure}

We selected five CAPTCHA-solving service retailers as our test target: 2Captcha\footnote{https://2captcha.com}, BestCaptchaSolver\footnote{https://bestcaptchasolver.com}, AntiCaptcha\footnote{https://anti-captcha.com}, DeathByCaptcha\footnote{https://www.deathbycaptcha.com}, and CapSolver\footnote{https://www.capsolver.com/}. They have relatively high ranks in Google's search results, support more CAPTCHA providers than others, and reveal more statistics details on their website. It is worth mentioning that 2Captcha, BestCaptchaSolver, AntiCaptcha, and DeathByCaptcha take 100\% human recognition as a feature worth advertising (Fig.~\ref{advertisement}). 2Captcha and DeathByCaptcha even specialized "100\% accuracy service", which works by sending the challenge to multiple workers and comparing the results. CapSolver, however, advertises that they solve CAPTCHAs using 100\% AI and machine learning methods.
\footnotetext{https://anti-captcha.com/}

Their prices for the selected four CAPTCHA providers are shown in Table~\ref{price}. Since the two types of Google reCAPTCHA V2 differ in labor required, some retailers will set different prices for them. The reason for price fluctuation for reCAPTCHA V3 is the workers' scores given by reCAPTCHA V3: workers with higher scores are more expensive. Overall, more difficult CAPTCHAs or those that take more time to solve will cost more to be solved. 

\begin{table}
\caption{Prices of CAPTCHA-solving services (\$/1,000 requests).}\label{price}
\centering
\begin{tabular}{|l|l|l|l|l|l|}
\hline
CAPTCHA type &  2Captcha & \makecell[l]{BestCaptcha\\Solver} & AntiCaptcha & \makecell[l]{DeathBy\\Captcha} & CapSolver\\
\hline
hCaptcha &  2.99 & 1.8 & 2 & 1.79 & 0.6\\
\hline
reCAPTCHA V2 &  1-2.99 & 2 & 0.95-2 & 2.89 & 0.8\\
\hline
reCAPTCHA V3 & 1.45-2.99 & 2 & 1-2 & 2.89 & 0.8\\
\hline
FunCaptcha & 2.99 & 2.5 & 3 & 3.99 & 2\\
\hline
GeeTest & 2.99 & 1.8 & 1.8 & 1.79 & 0.6\\
\hline
\makecell[l]{Text-based \\ CAPTCHA} & 1 & 0.6 & 0.7 & 1.39 & 0.4\\
\hline
\end{tabular}
\end{table}

2Captcha is the only retailer that recruits workers on the website. According to their online statistics, the wages are 0.26\$ per thousand correctly recognized text-based CAPTCHAs and 1\$ per thousand correctly solved image-based CAPTCHAs.

\section{Experiments}\label{experiment}
To look into the details of CAPTCHA-solving services, we ordered them from the retailers mentioned above and used them to solve popular CAPTCHAs. Since third-party CAPTCHAs don't vary with the website they are deployed on, we selected one website for each CAPTCHA provider. hCaptcha, Google reCAPTCHA, and GeeTest were tested on 2Captcha's CAPTCHA demo pages\footnote{https://2captcha.com/demo}. FunCaptcha was tested on Outlook's register page\footnote{https://outlook.live.com/owa/?nlp=1\&signup=1}. We also tested text-based CAPTCHA\footnote{https://www.mtcaptcha.com/\#mtcaptcha-demo}. We tested each CAPTCHA provider-solver pair once every five minutes for one day, and recorded the response time and result. Every attempt to test 2Captcha, DeathByCaptcha, and CapSolver against FunCaptcha got "Unsolvable" in response. Thus, these results are excluded when analyzing.

\subsection{Response time}
\begin{figure}[h]
\centering
\includegraphics[height=0.35\textheight]{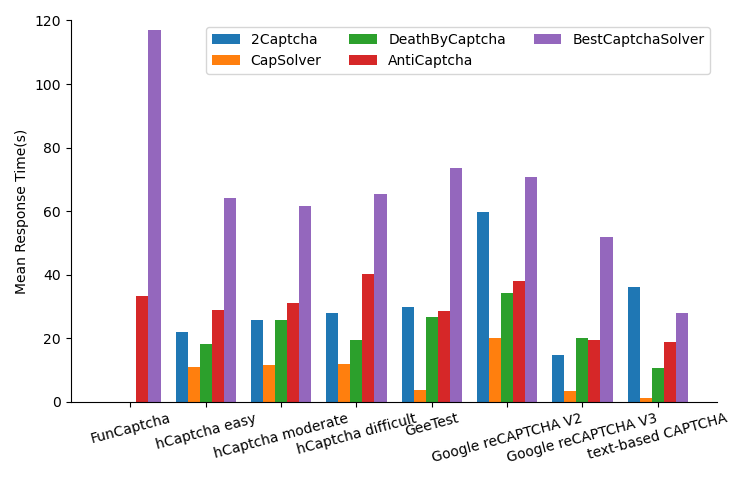}
\caption{Mean response time of different CAPTCHA-solving services when solving different CAPTCHAs}
\label{response_time}
\end{figure}

\begin{figure}[H]
\centering
\includegraphics[height=0.35\textheight]{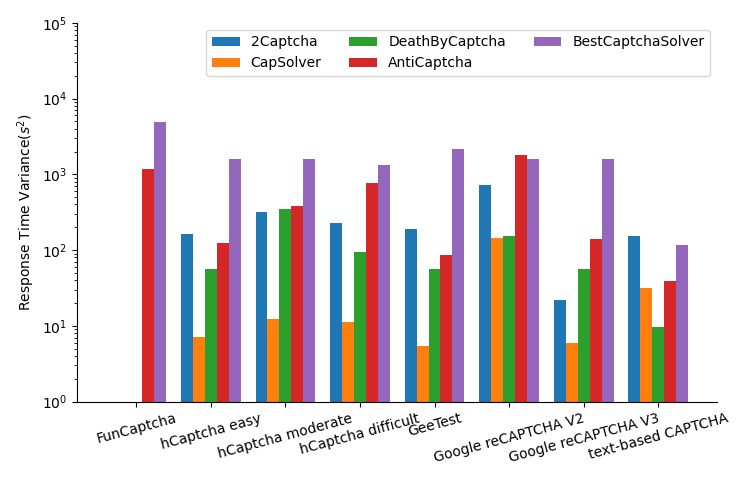}
\caption{Response time variance of different CAPTCHA-solving services when solving different CAPTCHAs}
\label{time_variance}
\end{figure}

Response time reflects the solving speed of CAPTCHA-solving services. Fig.~\ref{response_time} and Fig.~\ref{time_variance} shows the mean response time and the response time variance. Despite the mean response times of BestCaptchaSolver are abnormally high, we believe the reason is their high workload: their response time dropped to normal occasionally. From these experiments, we can conclude that the normal solving speed for image-based CAPTCHAs is 20 seconds to 40 seconds. We also observed that CapSolver has the lowest mean response time and response time variance against most CAPTCHAs. This proves that their "100\% AI and machine learning solutions" advertisement is not false. AntiCaptcha, on the contrary, has a relatively high mean response time, response time variance, and success rate, proving that they use 100\% human solvers.


\subsection{Success rate}
\begin{figure}[H]
\centering
\includegraphics[height=0.35\textheight]{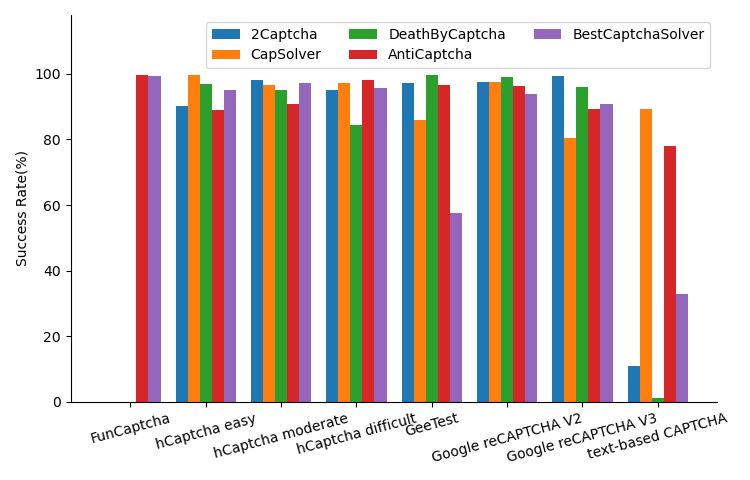}
\caption{Success rate of different CAPTCHA-solving services when solving different CAPTCHAs}
\label{success_rate}
\end{figure}

The success rate is a key argument for judging the quality of solving services and whether a CAPTCHA is broken. Fig.~\ref{success_rate} shows the success rate of different CAPTCHA-solving services when solving different CAPTCHAs. Overall, the results are similar to the previous studies' results: the solving services' success rates are relatively high. The success rates can reach 90\% in most cases. However, the situation is worse: the CAPTCHA providers are failing to stop automated solvers, let alone human solvers. The performance of CapSolver is worrying. Based on its mean response time, response time variance, and prices, we firmly believe it uses 100\% automated solvers, yet they have achieved success rates of over 80\% against most CAPTCHAs. This indicates that the latest hCaptcha, GeeTest, Google reCAPTCHA, and text-based CAPTCHA are already broken by automated solvers. Some success rates are lower, e.g., 2Captcha, DeathByCaptcha, and BestCaptchaSolver against text-based CAPTCHA. This does not indicate that text-based CAPTCHAs can effectively protect websites from CAPTCHA solvers since CapSolver's success rate against it is over 90\%, and CapSolver is using automated software solvers. We believe this is because they applied similar solvers but failed to recognize this specific text-based CAPTCHA.

\subsection{Difficulty}
\begin{figure}[h]
\centering
\includegraphics[height=0.27\textheight]{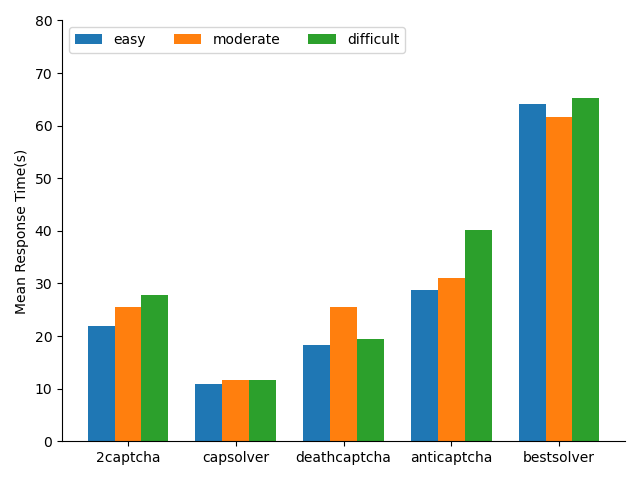}
\caption{Mean response time of different CAPTCHA-solving services when solving CAPTCHAs from hCaptcha with different difficulties}
\label{hcaptcha_response_time}
\end{figure}

\begin{figure}[h]
\centering
\includegraphics[height=0.27\textheight]{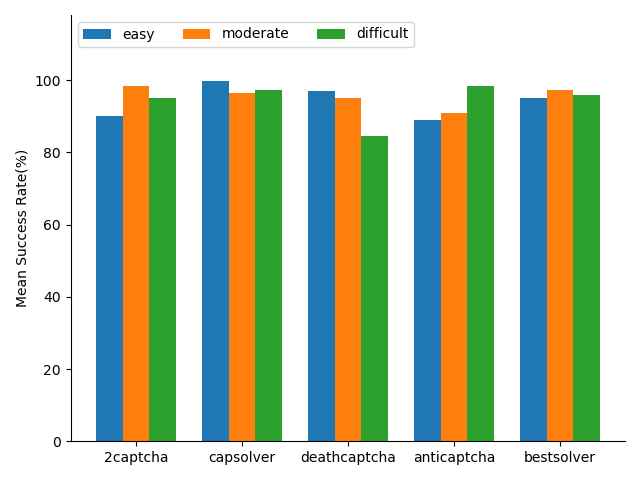}
\caption{Success rate of different CAPTCHA-solving services when solving CAPTCHAs from hCaptcha with different difficulties}
\label{hcaptcha_success_rate}
\end{figure}

Google reCAPTCHA, hCaptcha, and FunCaptcha claim to provide CAPTCHAs of different difficulties. The differences in CAPTCHA difficulty are the possibility of requiring the user to solve an explicit challenge and the difficulty of the challenge. Since hCaptcha is the only provider that allows the user to set difficulty manually, our comparative experiment is limited to hCaptcha. In Fig.~\ref{hcaptcha_response_time} and Fig.~\ref{hcaptcha_success_rate}, we compare the mean response time and success rate of different CAPTCHA-solving services when they are attempting to solve CAPTCHAs from hCaptcha with different difficulties. The result, however, proves that the difficulty of hCaptcha has little impact on both response time and success rate of the solving services.


\subsection{Workload}
\begin{figure}[H]
\centering
\begin{subfigure}[b]{0.95\textwidth}
\centering
\includegraphics[height=0.33\textheight]{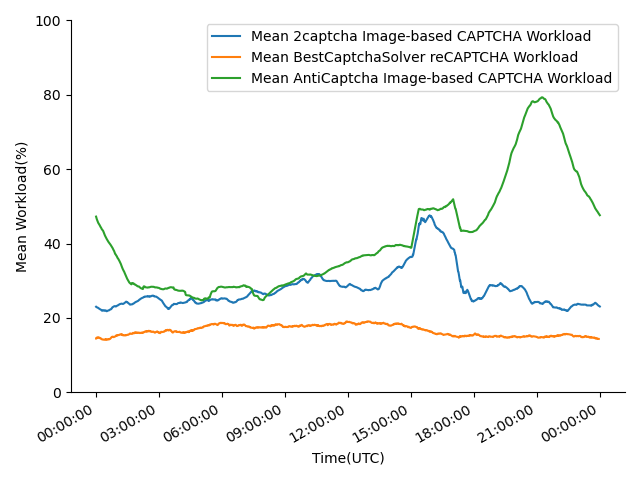}
\end{subfigure}
\begin{subfigure}[b]{0.95\textwidth}
\centering
\includegraphics[height=0.33\textheight]{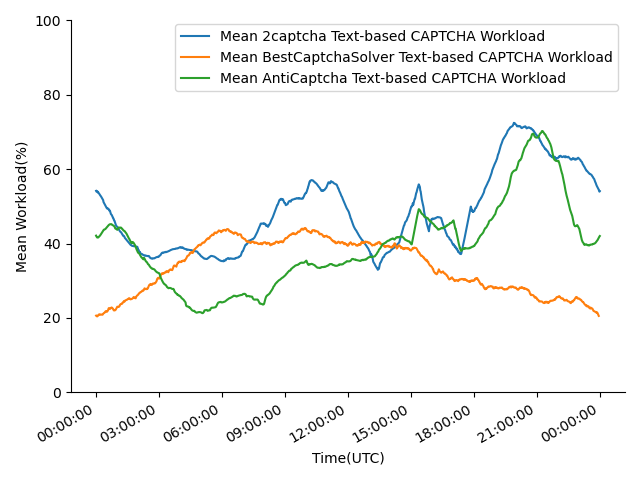}
\end{subfigure}
\begin{subfigure}[b]{0.95\textwidth}
\centering
\includegraphics[height=0.33\textheight]{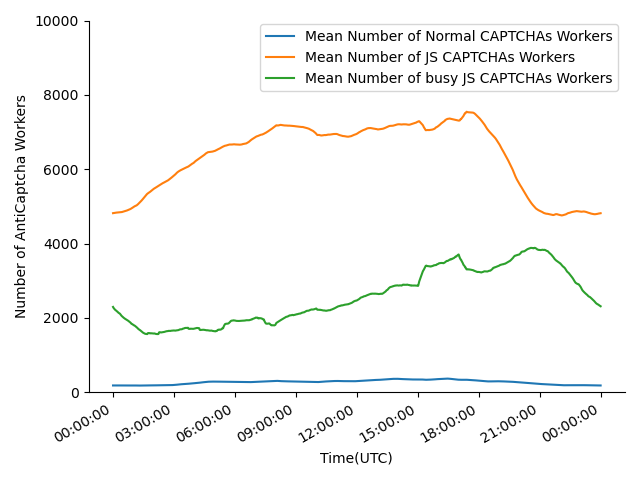}
\end{subfigure}
\caption{Reported workloads and worker numbers}
\label{workload}
\end{figure}

2Captcha, AntiCaptcha, and BestCaptchaSolver show their online statistics on their websites, mainly their real-time workload. When calculating the workloads, 2Captcha classifies the CAPTCHAs into normal CAPTCHAs and JS CAPTCHAs, BestCaptchaSolver only shows normal CAPTCHAs workload and reCAPTCHA workload, AntiCaptcha presents the number of busy workers and idle workers for each CAPTCHA they support. Since normal CAPTCHA usually refers to text-based CAPTCHAs, and the third-party CAPTCHAs mentioned above are JS-based, we will classify the workloads into normal/text-based CAPTCHAs workload and reCAPTCHA/JS-based CAPTCHAs workload. For data collected from AntiCaptcha, only the CAPTCHAs mentioned above are considered. We recorded these statistics once per minute for one day and drew mean workloads and AntiCaptcha worker numbers in 2 hours (Fig.~\ref{workload}). Although we didn't observe overload in experiments, there are obvious fluctuations and peaks in workloads. The number of AntiCaptcha workers between 18:00 UTC and 00:00 UTC is significantly lower than at other periods, corresponding to the peak after 18:00 UTC. We believe this is because 18:00 UTC to 00:00 UTC is the common sleeping time for low-income countries in south-east Asia. We can also observe that the number of workers solving JS CAPTCHAs is far more than normal CAPTCHAs since most websites use third-party CAPTCHAs.


\section{Conclusion}\label{conclusion}
CAPTCHA, a test meant to protect the Internet from underground businesses, ironically gave birth to another underground business: CAPTCHA farm. Both CAPTCHAs and CAPTCHA solvers have evolved in the past decade. This study presents the details of the latest third-party CAPTCHA providers, corresponding solving services, and the adversarial experiments between them. Based on the experiment results, we draw the following conclusions:
\newline \textbf{CAPTCHA providers cannot protect websites from human CAPTCHA solvers.} All selected popular third-party CAPTCHAs can be solved by at least two CAPTCHA-solving services with a high success rate.
\newline \textbf{CAPTCHA providers have relatively adequate human resources.} No overload was observed in the online statistics reported by CAPTCHA-solving service retailers. If overloaded, we expect the response time to be longer. Except for BestCaptchaSolver, other retailers kept the response time reasonable. Few failures due to timeout were recorded.
\newline \textbf{CAPTCHA providers are failing to stop automated solvers.} All selected popular third-party CAPTCHAs except FunCaptcha can be solved by CapSolver with a high success rate at a low price. 

CAPTCHAs were initially designed to distinguish humans and computer programs. However, the profit behind solving CAPTCHA has "transformed humans into computer programs." Although FunCaptcha can't stop human solvers, their hard AI problems are challenging to automated solvers and usually cost more to be solved, thus having an advantage compared to other CAPTCHAs. On the other hand, we believe that analyzing behaviors rather than focusing on the CAPTCHA challenge is a good start to distinguishing benign users and malicious attackers instead of distinguishing humans and computers. Still, CAPTCHA providers such as Google reCAPTCHA fail to fight against the latest CAPTCHA solvers and need more research.

\graphicspath{{./figs/}}
\bibliographystyle{splncs04}
\bibliography{main}

\end{document}